\documentclass[draft,tightenlines,nofootinbib,preprint,aps,eqsecnum,amsmath,amssymb]{revtex4}

\newcommand{\beq}{\begin{equation}}
\newcommand{\eeq}{\end{equation}}
\newcommand{\bea}{\begin{eqnarray}}
\newcommand{\eea}{\end{eqnarray}}

\newcommand{\sgn}{\epsilon}

\begin{document}

\title{A Relativistic Version of the Two-Level Atom
in the Rest-Frame Instant Form of Dynamics}

\author{David Alba}
\affiliation{Dipartimento di Fisica\\
Universita' di Firenze\\
Polo Scientifico, via Sansone 1\\
50019 Sesto Fiorentino, Italy\\
E-mail alba@fi.infn.it}
\author{Horace W. Crater}
\affiliation{The University of Tennessee Space Institute \\
Tullahoma, TN 37388 USA \\
E-mail: hcrater@utsi.edu}
\author{Luca Lusanna}
\affiliation{Sezione INFN di Firenze\\
Polo Scientifico\\
Via Sansone 1\\
50019 Sesto Fiorentino (FI), Italy\\
E-mail: lusanna@fi.infn.it}

\begin{abstract}

We define a relativistic version of the two-level atom, in which an
extended atom is replaced by a point particle carrying suitable
Grassmann variables for the description of the two-level structure
and of the electric dipole. After studying the isolated system "atom
plus the electro-magnetic field" in the electric-dipole
representation as a parametrized Minkowski theory, we give its
restriction to the inertial rest frame and the explicit form of the
Poincar\'e generators. After quantization we get a two-level atom
with a spin 1/2 electric dipole and the relativistic generalization
of the Hamiltonians of the Rabi and Jaynes-Cummings models.

\end{abstract}

\maketitle

\medskip

\today

\vfill\eject

\section{Introduction}

The Rabi model \cite{1a} and its rotating-wave approximation or
Jaynes-Cumming model \cite{1} of two-level atoms are  approximations
for the interaction of atoms with a quantized mode of an optical
cavity very useful in quantum optics and atomic physics. See
Ref.\cite{2} for an elementary introduction and the review in
Ref.\cite{3} for a detailed account of recent development.

\bigskip

This model like all atomic physics is an approximation to QED, in
which the atoms are described as non-relativistic particles in
quantum mechanics (QM) with a coupling to the electro-magnetic field
of order  $1/c$. For all the applications in which the energies
involved do not cross the threshold of pair production, this
description with a fixed number of particles is enough. Therefore
atomic physics is formulated in the absolute Euclidean 3-space and
use Newton absolute time, namely it is formulated in Galilei
space-time. The main drawback is that, due to the $1/c$ coupling to the
electro-magnetic field there is not a realization of the kinematical
Galilei group  connecting non-relativistic inertial frames. To get a
relativistic description we must reformulate the theory in Minkowski
space-time with a well defined realization of the kinematical
Poincar\'e group connecting relativistic inertial frames. This would
lead to {\it relativistic atomic physics} as the quantization of a
fixed number of classical relativistic charged scalar (or spinning)
particles interacting with the classical electro-magnetic field.

\medskip

In Refs.\cite{4,5} (based on the previous results of
Refs.\cite{6,7}) we gave a consistent relativistic formulation of
atomic physics with an explicit construction of the Poincar\'e
generators in the inertial rest-frame instant form of dynamics. In
these papers we considered N charged positive energy scalar
particles (with Grassmann-valued electric charges to regularize the
self-energies) interacting with the electro-magnetic field in the
radiation gauge in the framework of parametrized Minkowski theories
\cite{8}.\medskip

Parametrized Minkowski theories allows one to describe every isolated
system having a Lagrangian description in special relativity in both
inertial and non-inertial frames as shown in Refs. \cite{9,10}. This
formulation is based on a metrology-oriented description of
non-inertial frames obtained with the {\it 3+1 point of view} and
the use of observer-dependent Lorentz scalar radar 4-coordinates.
Let us give the world-line $x^{\mu}(\tau)$ of an arbitrary time-like
observer carrying a standard atomic clock: $\tau$ is an arbitrary
monotonically increasing function of the proper time of this clock.
Then we give an admissible 3+1 splitting of Minkowski space-time,
namely a nice foliation with space-like instantaneous 3-spaces
$\Sigma_{\tau}$: it is the mathematical idealization of a protocol
for clock synchronization (all the clocks in the points of
$\Sigma_{\tau}$ indicate the same time as the atomic clock of the
observer). On each 3-space $\Sigma_{\tau}$ we choose curvilinear
3-coordinates $\sigma^r$ having the observer as origin. These are
the radar 4-coordinates $\sigma^A = (\tau; \sigma^r)$. If $x^{\mu}
\mapsto \sigma^A(x)$ is the coordinate transformation from the
Cartesian 4-coordinates $x^{\mu}$ of a reference inertial observer
to radar coordinates, its inverse $\sigma^A \mapsto x^{\mu} =
z^{\mu}(\tau ,\sigma^r)$ defines the {\it embedding} functions
$z^{\mu}(\tau ,\sigma^r)$ describing the 3-spaces $\Sigma_{\tau}$ as an
embedded 3-manifold into Minkowski space-time. These embedding functions in turn induce a 4-metric
on $\Sigma_{\tau}$ by the following functional: 
${}^4g_{AB}(\tau ,\sigma^r) = [z^{\mu}_A\, \eta_{\mu\nu}\,
z^{\nu}_B](\tau ,\sigma^r)$, where $z^{\mu}_A(\tau, \sigma^r) =
\partial\, z^{\mu}(\tau, \sigma^r)/\partial\, \sigma^A$ are
tetrads (with inverse tetrads $z^A_{\mu}(\tau, \sigma^r)$) and
${}^4\eta_{\mu\nu} = \sgn\, (+---)$ is the flat metric ($\sgn = \pm
1$ according to either the particle physics $\sgn = 1$ or the
general relativity $\sgn = - 1$ convention). While the 4-vectors
$z^{\mu}_r(\tau ,\sigma^u)$ are tangent to $\Sigma_{\tau}$, so that
the unit normal $l^{\mu}(\tau ,\sigma^u)$ is proportional to
$\epsilon^{\mu}{}_{\alpha \beta\gamma}\, [z^{\alpha}_1\,
z^{\beta}_2\, z^{\gamma}_3](\tau ,\sigma^u)$, we have
$z^{\mu}_{\tau}(\tau ,\sigma^r) = [N\, l^{\mu} + N^r\,
z^{\mu}_r](\tau ,\sigma^r)$ ($N(\tau ,\sigma^r) = \sgn\,
[z^{\mu}_{\tau}\, l_{\mu}](\tau ,\sigma^r)$ and $N_r(\tau ,\sigma^r)
= - \sgn\, g_{\tau r}(\tau ,\sigma^r)$ are the lapse and shift
functions).\medskip

The foliation is nice and admissible if it satisfies the conditions
\footnote{These conditions imply that global {\it rigid} rotations
are forbidden in relativistic theories \cite{9,10}.}: \hfill\break
 1) $N(\tau ,\sigma^r) > 0$ in every point of
$\Sigma_{\tau}$ (the 3-spaces never intersect, avoiding the
coordinate singularity of Fermi coordinates);\hfill\break
 2) $\sgn\, {}^4g_{\tau\tau}(\tau ,\sigma^r) > 0$, so as to avoid the
 coordinate singularity of the rotating disk, and with the positive-definite 3-metric
${}^3g_{rs}(\tau ,\sigma^u) = - \sgn\, {}^4g_{rs}(\tau ,\sigma^u)$
having three positive eigenvalues (these are the M$\o$ller
conditions \cite{9});\hfill\break
 3) all the 3-spaces $\Sigma_{\tau}$ must tend to the same space-like
 hyper-plane at spatial infinity (so that there are always asymptotic inertial
observers to be identified with the fixed stars).\bigskip

In the description of isolated systems (particles, strings, fields,
fluids) admitting a Lagrangian formulation the matter variables are
replaced with new ones knowing the 3-spaces $\Sigma_{\tau}$. For a
relativistic particle with world-line $x^{\mu}(\tau )$ we must make
a choice of its energy sign: then it will be described by
3-coordinates $\eta^r(\tau )$ defined by the intersection of the
world-line with $\Sigma_{\tau}$: $x^{\mu}(\tau ) = z^{\mu}(\tau
,\eta^r(\tau ))$. As opposed to all the previous approaches to
relativistic mechanics, the dynamical configuration variables are
the 3-coordinates $\eta^r_i(\tau)$ and not the world-lines
$x^{\mu}_i(\tau)$ (to rebuild them in an arbitrary frame we need the
embedding defining that frame \cite{6}). For the electro-magnetic
field the potential ${\tilde A}_{\mu}(x^{\alpha})$ has to be
replaced with $A_A(\tau, \vec \sigma) = z^{\mu}_A(\tau, \vec \sigma)
\, {\tilde A}_{\mu}(z^{\alpha}(\tau, \vec \sigma))$, whose
associated field strength is $F_{AB}(\tau, \vec \sigma) =
\Big(\partial_A\, A_B -\partial_B\, A_A\Big)(\tau, \vec \sigma)$
(from now on we will use the vector notation $\vec \sigma$ for the
curvilinear 3-coordinates $\sigma^r$ for the sake of
simplicity).\medskip

Then the matter Lagrangian is coupled to an external gravitational
field and the external 4-metric is replaced with the 4-metric
$g_{AB}(\tau ,\vec \sigma)$ of an admissible 3+1 splitting of
Minkowski space-time. With this procedure we get a Lagrangian
depending on the given matter and on the embedding $z^{\mu}(\tau
,\vec \sigma)$, which is invariant under {\it frame-preserving
diffeomorphisms}. As a consequence, there are four first-class
constraints (an analogue of the super-Hamiltonian and super-momentum
constraints of canonical gravity) implying that the embeddings
$z^{\mu}(\tau ,\sigma^r)$ are {\it gauge variables}, so that all the
admissible non-inertial or inertial frames are gauge equivalent,
namely physics does {\it not} depend on the clock synchronization
convention and on the choice of the 3-coordinates $\sigma^r$: only
the appearances of phenomena change (not their order) by changing the notion of
instantaneous 3-space. Even if the gauge group is formed by the
frame-preserving diffeomorphisms, the matter energy-momentum tensor
allows the determination of the ten conserved Poincar\'e generators
$P^{\mu}$ and $J^{\mu\nu}$ (assumed finite) of every configuration
of the system.

 \medskip

If we restrict ourselves to inertial frames, we can define the {\it
inertial rest-frame instant form of dynamics for isolated systems}
by choosing the 3+1 splitting corresponding to the intrinsic
inertial rest frame of the isolated system centered on an inertial
observer: the instantaneous 3-spaces, named Wigner 3-space due to
the fact that the 3-vectors inside them are Wigner spin-1 3-vectors
\cite{7}, are orthogonal to the conserved 4-momentum $P^{\mu}$ of
the configuration. The embedding defining the inertial rest frame is

\beq
 z_{(W)}^{\mu}(\tau, \vec \sigma) = x_o^{\mu} +
 \epsilon^{\mu}_A(\vec h)\, \sigma^A,
 \label{1.1}
 \eeq

\noindent where $\epsilon^{\mu}_{\tau}(\vec h) = {{P^{\mu}}\over
{\sqrt{\sgn\, P^2}}} = h^{\mu} = \Big(\sqrt{1 + {\vec h}^2}; \vec
h\Big)$, $\epsilon^{\mu}_r(\vec h) = \Big(- h_r; \delta^i_r -
{{h^i\, h_r}\over {1 + \sqrt{1 + {\vec h}^2}}}\Big)$ (see
Ref.\cite{4} for the notations; we have $z^{\mu}_{(W) A}(\tau, \vec
\sigma) = \epsilon^{\mu}_A(\vec h)$). The tetrads
$\epsilon^{\mu}_A(\vec h)$ are the column of the standard Wigner
boost for time-like orbits; the inverse tetrads are
$\epsilon^A_{\mu}(\vec h)$

\bigskip

This framework allows the description of atomic physics \cite{4,5},
spinning particles \cite{11}, massless particles \cite{12}, open
Nambu string \cite{13}. Moreover it allows us to get a consistent
definition of  relativistic quantum mechanics and relativistic
entanglement in the inertial rest-frame \cite{14} with a preliminary
extension to non-inertial frames \cite{15,16}.
\bigskip

In Section 6 and 7 of Ref.\cite{5} we studied the electric-dipole
approximation and the transition to the relativistic electric-dipole
representation for a system of two charged positive-energy
particles. After the separation of the center of mass \footnote{See
Refs.\cite{4,5,9} for the problems connected with the relativistic
collective variables for an extended system, i.e. the canonical
non-covariant (Newton-Wigner) center of mass, the covariant
non-canonical Fokker-Pryce center of inertia and the non-covariant
non-canonical M$\o$ller center of energy. All of them collapse in
Newton center of mass in the non-relativistic limit.}, the relative
motion of the two particles gives rise to a relativistic bound state
at the quantum level whose levels simulate the levels of an atom.
\medskip

In this paper we will give a relativistic description of a two-level
atom with an electric dipole, in which we approximate the bound
state with its center of mass endowed with additional structures,
described by suitable Grassmann variables, which must generate only
a two-level structure (replacing the energy levels implied by the
Schroedinger equation for the relative motion) and a two-level
electric dipole (replacing the relative variable) after
quantization.\medskip

Therefore we get a particular monopole-dipole description of the
extended two-body system \footnote{See Section 5 of Ref.\cite{5} for
the standard multipolar expansion of the two-body system. It
includes the monopole (the center of mass) and the spin dipole (the
magnetic dipole), but not the electric dipole. Moreover it implies
all the levels of the two-body system after quantization and not
only two.}, in which the point-like atom has:\hfill\break

a) a generalized dipole description of the two levels;\hfill\break

b) an electric dipole, which after quantization can assume two
values like the spin of a Pauli particle;\hfill\break

c) and an optional spin dipole if the two-level atom is spinning
(moreover, if the two-level is charged instead of being neutral as
it is always assumed, the spin dipole would also behave as a
magnetic dipole as shown in Ref.\cite{11}).

\bigskip

We will give an action principle including also the dynamical
electro-magnetic field (so that we have an isolated system), whose
interaction with the point-like atom is given in the electric-dipole
representation (it contains an interaction term $\vec d \cdot \vec
E$) and not in the standard one (containing the interaction term
${{d \vec x(t)}\over {dt}} \cdot \vec A(t, \vec x(t))$).

\bigskip

The aim of the paper is not to contribute to the phenomenology of
atomic physics, but only to build an explicit realization of the
Poincar\'e generators of the relativistic two-level atom.

\bigskip

In order to facilitate the comparison of the various terms of our
relativistic version with its nonrelativistic counterpart, we present here
the nonrelativisitic Hamiltonian model for a point-like two-level atom
interacting with a single mode electromagnetic field%
\begin{eqnarray}
H &=&\frac{\vec{\kappa}^{2}}{2m}+\frac{\hbar }{2}\hat{\Omega}\sigma
_{3}+\hbar \omega \left( \hat{a}_{em}^{\dag }\hat{a}_{em}+\frac{1}{2}\right) 
\nonumber \\
&&+(c\vec{d}\cdot \mathcal{\vec{E}(\tau ))(\sigma }_{+}+\sigma _{-})(\hat{a}%
_{em}+\hat{a}_{em}^{\dag })  \label{0}
\end{eqnarray}%
It consists of the kinetic energy operator $\vec{\kappa}^{2}/2m$ of the
center of mass motion, the Hamiltonian $\hbar \omega \left( \hat{a}%
_{em}^{\dag }\hat{a}_{em}+\frac{1}{2}\right) $ of the free field and the
Hamiltonian $\frac{\hbar }{2}\hat{\Omega}\sigma _{3}$ corresponding to the
two-internal states of the atom. \ The final term corresponds to the
interaction between these degrees of freedom. \ Even though there are no
two-level atoms in nature, one can achieve a configuration in which only, in
effect, two-levels are involved by optical pumping. \ It is our aim to not
only obtain the relativistic quantum Dirac Hamiltonian corresponding to this
model but also an explicit realization of the other Poincare generators,
particularly the interaction dependent boost. \ In \cite{3} and \cite{17}
the  details are presented of a microsopic derivation of Eq. (\ref{0}) for a
hydrogen atom in a time-varying electro-magnetic field. \ We will not be
presenting the  microscopic derivation of the relativisitic version\footnote{%
In \cite{10} we presented the classical version of this microscopic
derivation including the presentation of the Poincaire generators.}. \
Rather we will be arriving at our goal by first constructing the
(pseudo)-classical Lagrangian description of a point-like two-level atom as
a parametrized Minkowski theory. \ This is done in Section II. \ 

In Section III we study the constraints present in the Hamiltonian
description and then we give the restriction of the isolated system
to the inertial rest frame and the explicit form of the Poincar\'e
generators.

In Section IV we show how the quantization of the Grassmann
variables and of the constraints implies the two-level structure, a
two-level electric dipole and the Hamiltonian of the Rabi model.

Some final comments are done in the Conclusions.

\vfill\eject

\section{The Lagrangian Description of the Two-Level Atom in a
Dynamical Electro-Magnetic Field}

Let us consider the isolated system of a two-level atom interacting
with a dynamical electro-magnetic field in the framework of
parametrized Minkowski theories..
\medskip

The point-like two-level atom is described by the 3-coordinates
$\eta^r(\tau)$, so that its world-line is given by $x^{\mu}(\tau) =
z^{\mu}(\tau, \eta^r(\tau))$ as said in the Introduction. This is
the monopole description of an extended atom.

\medskip

The monopole carries two pairs of complex Grassmann variables
$\alpha(\tau)$, $\alpha^*(\tau)$, $\beta(\tau)$, $\beta^*(\tau)$
needed for the description of a {\it structure with only two energy
levels}. They satisfy $\alpha^2(\tau) = (\alpha^*(\tau))^2 =
\alpha(\tau)\, \alpha^*(\tau) + \alpha^*(\tau)\, \alpha(\tau)= 0$,
$\beta^2(\tau) = (\beta^*(\tau))^2 = \beta(\tau)\, \beta^*(\tau) +
\beta^*(\tau)\, \beta(\tau)= 0$, $\alpha(\tau)\, \beta(\tau) -
\beta(\tau)\, \alpha(\tau)= \alpha(\tau)\, \beta^*(\tau) -
\beta^*(\tau)\, \alpha(\tau) = \alpha^*(\tau)\, \beta(\tau) -
\beta(\tau)\, \alpha^*(\tau) = \alpha^*(\tau)\, \beta^*(\tau) -
\beta^*(\tau)\, \alpha^*(\tau) = 0$. As shown in Ref.\cite{11} the
quantization of each pair of complex Grassmann variables generates a
two-level Fermi oscillator. These variables thus underly the quantum operators 
corresponding to the Pauli-matrices in Eq. (\ref{0})

\medskip

To describe the {\it electric dipole} carried by the atom we use a
real Grassmann 4-vector $\xi^{\mu}(\tau)$, $\xi^{\mu}(\tau)\,
\xi^{\nu}(\tau) + \xi^{\nu}(\tau)\, \xi^{\mu}(\tau) = 0$, commuting
with the Grassmann variables describing the two-level structure. To
describe a two-level structure after quantization (like a spin 1/2
Pauli particle) it must satisfy the constraint $P_{\mu}\, \xi^{\mu}
\approx 0$, where $P^{\mu}$ is the total conserved 4-momentum of the
isolated system. Since, as shown in Ref.\cite{11}, it is too
difficult to include this condition with a Lagrange multiplier in
the Lagrangian, we will give a Lagrangian not implying this
condition. Instead we will add it by hand in the resulting
Hamiltonian description. Then conceivably one could try to derive
the correct Lagrangian by means of the inverse Legendre
transformation. The variables $\xi^{\mu}(\tau)$ allow us to build a
spin-like tensor $S^{\mu\nu}(\tau) = - i\, \xi^{\mu}(\tau)\,
\xi^{\nu}(\tau)$ satisfying $P_{\mu}\, S^{\mu\nu}(\tau) \approx 0$.
In the Wigner 3-spaces of the rest frame the constraint $P_{\mu}\,
\xi^{\mu} \approx 0$ implies that  only the following three
Grassmann variables survive

 \beq
 \xi^r_{\perp}(\tau)\, {\buildrel {def}\over =}\, \epsilon^r_{\mu}(\vec h)\,
 \xi^{\mu}(\tau).
 \label{2.1}
 \eeq

\noindent Then the rest-frame spin-like tensor ${\bar S}^{AB}(\tau) =
\epsilon^A_{\mu}(\vec h)\, \epsilon^B_{\nu}(\vec h)\,
S^{\mu\nu}(\tau)$ satisfies ${\bar S}^{\tau B}(\tau) \approx 0$ and
${\bar S}^r(\tau) = {1\over 2}\, \epsilon^{ruv}\, {\bar
S}^{uv}(\tau) = - {i\over 2}\, \epsilon^{ruv}\,
\xi_{\perp}^u(\tau)\, \xi_{\perp}^v(\tau)$. As shown in
Ref.\cite{11} the quantization of the three real Grassmann variables
$\xi_{\perp}^r$ generates the algebra of Pauli matrices
($\xi_{\perp}^r \mapsto \sqrt{{{\hbar}\over 2}}\, \sigma^r$)
describing the $({1\over 2}, 0)$ representation of SL(2, C).
They will be used in the construction of the dipole operator $\vec d$ in Eq. (\ref{0})

\bigskip

In this paper we consider only the {\it electric dipole}. If the
two-level atom is charged, we could also introduce the {\it magnetic
dipole} by adding a spin dipole described by a real Grassmann
4-vector $\xi^{\mu}_Q(\tau)$ satisfying $P_{\mu}\, \xi^{\mu}_Q(\tau)
\approx 0$ as done in Ref.\cite{11} to describe a positive
energy spinning particle giving rise to a positive energy spin 1/2
particle at the quantum level. In this case we need also a
Grassmann-valued electric charge $Q(\tau) = \theta^*(\tau)\,
\theta(\tau)$ described by a pair of complex Grassmann variables (it
turns out to be a constant of motion, ${{d Q(\tau)}\over {d \tau}} =
0$ ). Again each type of Grassmann variable is commuting with the
Grassmann variables of a different type.
\medskip

In Ref.\cite{11} the mass term in the Lagrangian for the spinning
particle was

\bea
 &&\sqrt{m^2\, c^2 +  Q\, S_Q^{\mu\nu}(\tau)\,
 F_{\mu\nu}(z(\tau, \vec \sigma))} =\nonumber \\
 &&= \sqrt{m^2\, c^2 - i\, Q\,
 \xi_Q^{\mu}(\tau)\, \xi^{\nu}_Q\, z^A_{\mu}(\tau, \vec
 \sigma)\, z^B_{\nu}(\tau, \vec \sigma)\, F_{AB}(\tau, \vec \sigma)}
 \approx\nonumber \\
 &&\approx \sqrt{m^2\, c^2 - 2\, Q\, {\vec {\bar S}}_Q \cdot
 {\vec B}(\tau, \vec \sigma)} =\nonumber \\
 &&= m c - Q\, {\vec {\bar S}}_Q
 \cdot {\vec B}(\tau, \vec \sigma),
 \label{2.2}
 \eea

\noindent where we used the implications of the constraint
$P_{\mu}\, \xi^{\mu}_Q(\tau) \approx 0$ on the spin tensor
$S_Q^{\mu\nu} = - i\, \xi_Q^{\mu}(\tau)\, \xi_Q^{\nu}(\tau)$ to show
the emergence of the  coupling of the spin dipole with the magnetic
field.

\bigskip

By analogy we get the coupling of the electric dipole, described by the
Grassmann variables $\xi^{\mu}(\tau)$, to the electric field by modifying
Eq.(\ref{2.2}) through the introduction of a coupling of the associated
spin-like term  $S^{\mu\nu}_{\xi}(\tau) = - i\, \xi^{\mu}(\tau)\,
\xi^{\nu}(\tau)$ with the dual $F^*_{\mu\nu} = {1\over 2}\,
\epsilon_{\mu\nu\alpha\beta}\, F^{\alpha\beta}$ of the field
strength: now $S^{\mu\nu}\, F^*_{\mu\nu} = - {i\over 2}\,
\xi^{\mu}\, \xi^{\nu}\, \epsilon_{\mu\nu}{}^{\alpha\beta}\,
z_{\alpha}^A\, z_{\beta}^B\, F_{AB}$ will become ${\vec {\bar S}}
\cdot {\vec E}$ which will correspond to the electric-dipole coupling.

\bigskip

The previous discussion and the form of the Hamiltonian given in Eq. (\ref{0}) of the
Jaynes-Cummings model suggests replacing the mass $mc$ with the
following expression

\bea
 m^*(\tau, \vec \eta(\tau))\, c &=& mc + \Omega\, \beta^*(\tau)\, \beta(\tau)
 +\nonumber \\
 &+& {i\over 2}\, d\, (\beta^*(\tau)\, \alpha(\tau) + \alpha^*(\tau)\,
 \beta(\tau))\, \xi^{\mu}(\tau)\, \xi^{\nu}(\tau)\,
 \epsilon_{\mu\nu}{}^{\alpha\beta}\, \Big(z_{\alpha}^A(\tau, \vec
 \sigma)\, z_{\beta}^B(\tau, \vec \sigma)\nonumber \\
 &&\times F_{AB}(\tau, \vec
 \sigma)\Big){|}_{\vec \sigma = \vec \eta(\tau)}.\nonumber \\
 &&{}
 \label{2.3}
 \eea

\noindent Here $\Omega$ is the energy difference between the two
levels and $d$ is the coupling constant between the electric dipole
and the electric field.

\bigskip

Therefore in the framework of parametrized Minkowski theories our
isolated system is described:\hfill\break

a) by the 3-coordinates $\eta^r(\tau)$ of the point-like
atom;\hfill\break

b) by two pairs of complex Grassmann variables $\alpha(\tau)$ and
$\beta(\tau)$ describing the levels of the atom;\hfill\break

c) by a real Grassmann 4-vector $\xi^{\mu}(\tau)$ (to be reduced to
a Wigner spin-1 3-vector $\xi_{\perp}^r(\tau)$ at the Hamiltonian
level with the addition of the constraint $P_{\mu}\, \xi^{\mu}(\tau)
\approx 0$) to describe the electric dipole;\hfill\break

d) by the electro-magnetic potential $A_A(\tau, \vec \sigma)$
defined in the Introduction;\hfill\break

e) by the embedding $z^{\mu}(\tau, \vec \sigma)$ describing the 3+1
splitting, which will be restricted to the embedding of the Wigner
3-spaces of the inertial rest frame at the Hamiltonian level at the
end.

\bigskip

The Lagrangian of the system, describing it as a parametrized
Minkowski theory, is ($\dot a(\tau) = {{d a(\tau)}\over {d \tau}}$;
$m^*(\tau, \vec \sigma)$ given in Eq.(\ref{2.3}))

\bea
 L(\tau) &=& \int d^3\sigma\, {\cal L}(\tau, \vec \sigma),
 \nonumber \\
 &&{}\nonumber \\
 {\cal L}(\tau, \vec \sigma) &=&
 \delta^3(\vec \sigma - \vec \eta(\tau))\nonumber \\
 &&\Big[ {i\over 2}\, \Big(\xi_{\mu}(\tau)\,
 {\dot \xi}^{\mu}(\tau) + \beta^*(\tau)\, \dot \beta(\tau) -
 {\dot \beta}^*(\tau)\, \beta(\tau) + \alpha^*(\tau)\, \dot \alpha(\tau)
 - {\dot \alpha}^*(\tau)\, \alpha(\tau)\Big) -\nonumber \\
 &-& m^*(\tau, \vec \sigma)\, c\, \sqrt{{}^4g_{\tau\tau} + 2\,
 {}^4g_{\tau r}\, {\dot \eta}^r(\tau) + {}^4g_{rs}\,
 {\dot \eta}^r(\tau)\, {\dot \eta}^s(\tau)}(\tau, \vec \sigma)
 +\nonumber \\
 &+& \lambda_1(\tau)\, \Big(\alpha^*(\tau)\, \alpha(\tau) +
 \beta^*(\tau)\, \beta(\tau)\Big)\Big] -\nonumber \\
 &-& {1\over 4}\, \Big(\sqrt{|{}^4g|}\, {}^4g^{AC}\, {}^4g^{BD}\,
 F_{AB}\, F_{CD}\Big)(\tau, \vec \sigma).
 \label{2.4}
 \eea

\medskip

Its construction will ultimately allow us to obtain expressions for the Poincaire generators of Eq. (\ref{3.17}).
The first term is the kinetic term for the Grassmann variables. The
Lagrange multiplier $\lambda_1(\tau)$ is needed to get a constraint
on the Grassmann variables which reduces the levels of the atom from
four to two and is considered as a configuration variable.\medskip

Let us remark that the analogous Lagrangian for the spinning
particle of Ref. \cite{11}: 1) does not contains the variables
$\alpha(\tau)$ and $\beta(\tau)$ (but has the Grassmann variables
for the electric charge $Q = \theta^*(\tau)\, \theta(\tau)$); 2) has
$mc + \Omega\, \beta^*(\tau)\, \beta(\tau)$ replaced by the ordinary
mass term $m c$; 3) has the electric-dipole term ${i\over 2}\, d\,
(\beta^*(\tau)\, \alpha(\tau) + \alpha^*(\tau)\,
 \beta(\tau))\, \xi^{\mu}(\tau)\, \xi^{\nu}(\tau)\,
 \epsilon_{\mu\nu}{}^{\alpha\beta}$ replaced by the spin-dipole term
$- {i\over {2 m c}}\, Q\, \xi^{\alpha}_Q(\tau)\,
 \xi_Q^{\beta}(\tau) $; 4) has the minimal coupling term
$\delta^3(\vec \sigma - \vec \eta(\tau))\, Q\, \Big(A_{\tau}(\tau,
\vec \sigma) + {\dot \eta}^r(\tau)\, A_r(\tau, \vec \sigma)\Big)$
not being in the electric dipole representation. If we would
consider a charged and spinning two-level atom, instead of the
standard neutral one, we should add terms like in items 1), 3) and
4) to the Lagrangian (\ref{2.4}).

\bigskip

The canonical momenta associated with the Grassmann variables and with
the Lagrangian multiplier are

\bea
 \pi^\mu_\xi(\tau) &=& \frac{i}{2}\, \xi^\mu(\tau),\nonumber \\
 \pi_\alpha(\tau) &=&- \frac{i}{2}\, \alpha^*(\tau),\qquad
 \pi^*_\alpha(\tau) = + \frac{i}{2}\,\alpha(\tau),\nonumber \\
 \pi_\beta(\tau) &=&- \frac{i}{2}\, \beta^*(\tau),\qquad
 \pi^*_\beta(\tau) = + \frac{i}{2}\, \beta(\tau),\nonumber \\
 \pi_{\lambda_1}(\tau) &=& 0.
 \label{2.5}
 \eea

\medskip

If we introduce the following notation for the electric dipole

\beq
 A^{\alpha\beta}(\tau) = {i\over 2}\, d\,
 \Big(\beta^*(\tau)\, \alpha(\tau) + \alpha^*(\tau)\,
 \beta(\tau)\Big)\, \xi^{\mu}(\tau)\, \xi^{\nu}(\tau)\,
 \epsilon_{\mu\nu}{}^{\alpha\beta},
 \label{2.6}
 \eeq

\noindent we have that the terms bilinear in this quantity vanish
because we have $\Big(\beta^*(\tau)\, \alpha(\tau) +
\alpha^*(\tau)\, \beta(\tau)\Big)^2 = 0$ due to the properties of
the Grassmann variables.\medskip

Then the canonical momenta associated with the atom 3-position and with
the electro-magnetic field have the following forms

\bea
 \kappa_r(\tau) &=&- \frac{\partial L(\tau)}{\partial\,\dot{\eta}^r} =
 \Big(m c + \Omega\, \beta^*(\tau)\, \beta(\tau) +\nonumber \\
 &+&A^{\alpha\beta}(\tau)\, z^A_\alpha(\tau, \vec \eta(\tau))\, z^B_\beta(\tau,
 \vec \eta(\tau))\,F_{AB}(\tau, \vec \eta(\tau))\Big)\nonumber \\
 &&\frac{^4g_{\tau r}(\tau, \vec \eta(\tau))
 + {}^4g_{rs}(\tau, \vec \eta(\tau))\, \dot{\eta}^r(\tau)}
 {\sqrt{{}^4g_{\tau\tau}(\tau, \vec \eta(\tau)) + 2\, {}^4g_{\tau u}(\tau,
 \vec \eta(\tau))\, \dot{\eta}^u(\tau) + {}^4g_{uv}(\tau, \vec \eta(\tau))\,
 \dot{\eta}^u(\tau)\, \dot{\eta}^v(\tau)}},\nonumber \\
 &&{}\nonumber \\
 \pi^\tau(\tau, \vec \sigma) &=& \frac{\partial {\cal L}(\tau, \vec \sigma)}
 {\partial(\partial_\tau A_\tau)} = 0,\nonumber \\
 &&{}\nonumber \\
 \pi^r(\tau, \vec \sigma) &=&\frac{\partial {\cal L(\tau, \vec \sigma)}}
 {\partial(\partial_\tau A_r)} =
 - \Big[\frac{\gamma}{\sqrt{{}^4g}}\, \gamma^{rs}\, \Big(F_{\tau s} - {}^4g_{\tau
 v}\, \gamma^{uv}\, F_{us}\Big)\Big](\tau, \vec \sigma) -\nonumber \\
 &-& \delta^3(\vec \sigma - \vec \eta(\tau))\, \frac{2\,
 \Big(m c + \Omega\, \beta^*(\tau)\, \beta(\tau)\Big)\,
 A^{\alpha\beta}(\tau)\, \Big(l_\alpha\, z_{s\beta}\, \gamma^{sr}\Big)(\tau,
 \vec \sigma)}{\sqrt{m^2\, c^2 + 2\, mc\, \Omega\, \beta^*(\tau)\, \beta(\tau)
 + \gamma^{uv}(\tau, \vec \sigma)\, \kappa_u(\tau)\, \kappa_v(\tau)}},
 \nonumber \\
 &&{}
 \label{2.7}
 \eea

\noindent where $\gamma^{rs}(\tau, \vec \sigma)$ is the inverse of
the positive-signature 3-metric ${}^3g_{rs}(\tau, \vec \sigma) = -
\sgn\, {}^4g_{rs}(\tau, \vec \sigma)$, $\gamma(\tau, \vec \sigma) =
det\, {}^3g_{rs}(\tau, \vec \sigma)$, ${}^4g(\tau, \vec \sigma) =
|det\, {}^4g_{AB}(\tau, \vec \sigma)|$ and $l^{\alpha}(\tau, \vec
\sigma)$ is the unit normal to the 3-space $\Sigma_{\tau}$ in the
point with 3-coordinates $\vec \sigma$.

\medskip

Finally the canonical momentum conjugate to the embedding
$z^{\mu}(\tau, \vec \sigma)$ is

\bea
 \rho_\mu(\tau, \vec \sigma) &=& - \frac{\partial {\cal L}(\tau, \vec \sigma)}
 {\partial z^\mu_\tau} = \delta^3(\vec{\sigma} -
 \vec{\eta}(\tau))\nonumber \\
 &&\Big[\,\,\,\Big(m c + \Omega\, \beta^*(\tau)\, \beta(\tau) + A^{\alpha\beta}(\tau)\,
 z^A_\alpha(\tau, \vec \sigma)\, z^B_\beta(\tau, \vec \sigma)\,
 F_{AB}(\tau, \vec \sigma)\Big)\nonumber \\
 &&\frac{z_{\tau\mu}(\tau, \vec \sigma) +
  z_{r\mu}(\tau, \vec \sigma)\, \dot{\eta}^r(\tau)}
  {\sqrt{{}^4g_{\tau\tau}(\tau, \vec \sigma) +
 2\, {}^4g_{\tau u}(\tau, \vec \sigma)\, \dot{\eta}^u(\tau) +
 {}^4g_{uv}(\tau, \vec \sigma)\, \dot{\eta}^u(\tau)\, \dot{\eta}^v(\tau)}}\,
 +\nonumber \\
 &&{}\nonumber \\
 &-&2\, A^{\rho\sigma}(\tau)\, \Big[
 z_{\tau\mu}\, ({}^4g^{A\tau}\, {}^4g^{\tau C}\, {}^4g^{BD} +
 {}^4g^{AC}\, {}^4g^{B\tau}\, {}^4g^{\tau D}) +\nonumber \\
 &+& z_{r\mu}\, ({}^4g^{Ar}\, {}^4g^{\tau C} + {}^4g^{A\tau}\, {}^4g^{rC})\,
 {}^4g^{BD}\Big](\tau, \vec \sigma) \times \nonumber \\
 &&{}\nonumber \\
 &&\times\, \Big(z_{C\rho}\, z_{D\nu}\, F_{AB}\Big)(\tau, \vec \sigma)\,
 \sqrt{{}^4g_{\tau\tau}(\tau, \vec \sigma) + 2\, {}^4g_{\tau u}(\tau, \vec \sigma)\,
 \dot{\eta}^u(\tau) + {}^4g_{uv}(\tau, \vec \sigma)\, \dot{\eta}^u(\tau)\,
 \dot{\eta}^v(\tau)}\, +\nonumber \\
 &&{}\nonumber \\
 &-&2\, A_{\mu\rho}(\tau)\, \Big({}^4g^{A\tau}\,
 {}^4g^{BD}\, z^\rho_D\, F_{AB}\Big)(\tau, \vec \sigma)\nonumber \\
 &&\sqrt{{}^4g_{\tau\tau}(\tau, \vec \sigma) + 2\, {}^4g_{\tau u}(\tau,
 \vec \sigma)\, \dot{\eta}^u(\tau) + {}^4g_{uv}(\tau, \vec \sigma)\,
 \dot{\eta}^u(\tau)\, \dot{\eta}^v(\tau)}\,\,\, \Big] +\nonumber \\
 &&{}\nonumber \\
 &+& \frac{\sqrt{{}^4g}}{4}\, \Big[\, {}^4g^{\tau E}\, z_{E\mu}\, {}^4g^{AC}\,
 {}^4g^{BD}\, F_{AB}\, F_{CD} - 2\, z_{\tau\mu}\, ({}^4g^{A\tau}\, {}^4g^{\tau C}\,
 {}^4g^{BD} + {}^4g^{AC}\, {}^4g^{B\tau}\, {}^4g^{\tau D})\, F_{AB}\,F_{CD}
 +\nonumber \\
 &-&2\, z_{r\mu}\, ({}^4g^{Ar}\, {}^4g^{\tau C} + {}^4g^{A\tau}\, {}^4g^{rC})\,
 {}^4g^{BD}\, F_{AB}\, F_{CD}\,\Big](\tau, \vec \sigma).
 \label{2.8}
 \eea

\medskip

The canonical Hamiltonian is

\beq
 H_c = - \lambda_1(\tau)\, \Big(\alpha^*(\tau)\, \alpha(\tau) +
 \beta^*(\tau)\, \beta(\tau)\Big) - \int d^3\sigma\, A_{\tau}(\tau, \vec
 \sigma)\, \partial_r\, \pi^r(\tau, \vec \sigma).
 \label{2.9}
 \eeq

\noindent while the conserved Poincar\'e generators are (see
Ref.\cite{8,11})

\bea
 P^\mu &=& \int d^3\sigma\, \rho^\mu(\tau,\vec{\sigma}),\nonumber \\
 &&{}\nonumber \\
 J^{\mu\nu}&=& \int d^3\sigma\, \Big[z^\mu(\tau,\vec{\sigma})\,\rho^\nu(\tau,\vec{\sigma})
 - z^\nu(\tau,\vec{\sigma})\, \rho^\mu(\tau,\vec{\sigma})\Big] + i\, \xi^\mu(\tau)\,
 \xi^\nu(\tau).
 \label{2.10}
 \eea

\vfill\eject

\section{The Hamiltonian Description}

Let us now describe the primary Dirac constraints of the model.

\medskip

The Grassmann momenta (\ref{2.5}) imply the following second class
constraints

\bea
 \chi^\mu(\tau) &=& \pi_\xi^\mu(\tau) - \frac{i}{2}\, \xi^\mu(\tau) \approx 0, \nonumber \\
 &&{}\nonumber \\
 \chi_{(\beta)}(\tau) &=& \pi_\beta(\tau) + \frac{i}{2}\, \beta^*(\tau) \approx 0,
 \qquad \chi_{(\beta)}^*(\tau) = \pi_\beta^*(\tau) - \frac{i}{2}\, \beta(\tau)
 \approx 0,\nonumber \\
 &&{}\nonumber \\
 \chi_{(\alpha)}(\tau) &=& \pi_\alpha(\tau) + \frac{i}{2}\, \alpha^*(\tau) \approx 0,\qquad
 \chi_{(\alpha)}^*(\tau) = \pi_\alpha^*(\tau) - \frac{i}{2}\, \alpha(\tau) \approx
 0.
 \label{3.1}
 \eea

The Poisson brackets of the Grassmann variables are $\{
\xi^{\mu}(\tau), \pi_{\xi}^{\nu}(\tau) \} = - \eta^{\mu\nu}$, $\{
\alpha(\tau), \pi_{\alpha}(\tau) \} = \{ \alpha^*(\tau),
\pi^*_{\alpha}(\tau) \} = \{ \beta(\tau), \pi_{\beta}(\tau) \} = \{
\beta^*(\tau), \pi^*_{\beta}(\tau) \} = - 1$.\medskip

The other primary constraints are

\bea
 \pi_{\lambda_1}(\tau) &\approx& 0,\nonumber \\
 \pi^{\tau}(\tau, \vec \sigma) &\approx& 0,\nonumber \\
 {\cal H}_{\mu}(\tau, \vec \sigma) &\approx& 0,
 \label{3.2}
 \eea

\noindent where, following the methods of Ref.\cite{11} and using
the results given after Eq.(\ref{2.6}), one has the following final
expression for the constraints ${\cal H}_{\mu}(\tau, \vec \sigma)
\approx 0$ deriving from  Eqs.(\ref{2.8}) (they imply that the
embeddings are gauge variables or equivalently, they provide a 
four-vector continuum set of first class constraints associated 
with the invariance of the action under $\tau$ and $\vec\sigma$ parameter changes).

\bea
 {\cal H}_\mu(\tau,\vec{\sigma})&=&\rho_\mu(\tau,\vec{\sigma}) -
  z_{r\mu}(\tau,\vec{\sigma})\, \gamma^{rs}(\tau,\vec{\sigma})\,
 \Big[\,\,\delta^3(\vec{\sigma} - \vec{\eta}(\tau))\, \kappa_r(\tau) +
 F_{ru}(\tau,\vec{\sigma})\, \pi^u(\tau,\vec{\sigma})\,\,\Big] -
 \nonumber \\
 &&{}\nonumber \\
 &-&l_\mu(\tau,\vec{\sigma})\, \Big[\,\,\delta^3(\vec{\sigma} - \vec{\eta}(\tau))\,
 \Big(\sqrt{m^2\, c^2 + 2\, m c\, \Omega\, \beta^*(\tau)\,
 \beta(\tau) + \gamma^{rs}(\tau,\vec{\sigma})\, \kappa_r(\tau)\,
 \kappa_s(\tau)} +\nonumber \\
 &&{}\nonumber \\
 &-&\frac{2\, (m c + \Omega\, \beta^*(\tau)\, \beta(\tau))}{\sqrt{\gamma(\tau,\vec{\sigma})}}\,
 \frac{A^{\mu\nu}(\tau)\, \Big(l_\mu\, z_{s\nu}\, \pi^s\Big)(\tau,\vec{\sigma})}
 {\sqrt{m^2\, c^2 + 2\, mc\, \Omega\, \beta^*(\tau)\, \beta(\tau) +
 \gamma^{rs}(\tau,\vec{\sigma})\, \kappa_r(\tau)\,
 \kappa_s(\tau)}}+\nonumber \\
 &+& \Big( m c + \Omega\, \beta^*(\tau)\, \beta(\tau) \Big)\,
 \frac{A^{\mu\nu}(\tau)\, \Big(z_{u\mu}\, z_{v\nu}\, \gamma^{ur}\,
 \gamma^{vs}\, F_{rs}\Big)(\tau,\vec{\sigma})}{\sqrt{m^2\, c^2 +
 \Omega\, \beta^*(\tau)\, \beta(\tau)
 + \gamma^{rs}(\tau,\vec{\sigma})\, \kappa_r(\tau)\, \kappa_s(\tau)}} \Big)
 +\nonumber \\
 &&{}\nonumber \\
 &+&\Big(- \frac{1}{2\sqrt{\gamma}}\, {}^4g_{rs}\, \pi^r\, \pi^s + \frac{\sqrt{\gamma}}{4}\,
 \gamma^{rs}\, \gamma^{uv}\, F_{ru}\, F_{sv}\Big)(\tau,\vec{\sigma})\,\,\Big]
 =\nonumber \\
 &=& \rho_{\mu}(\tau, \vec \sigma) - \Big(\sqrt{\gamma}\, \Big[l_{\mu}\,
 T_{\perp\perp} - z_{r\mu}\, \gamma^{rs}\, T_{\perp s}\Big]\Big)(\tau, \vec \sigma)
 \approx 0,\nonumber \\
 &&{}
 \label{3.3}
 \eea

\noindent where $T_{\perp\perp} = l_{\mu}\, l_{\nu}\, T^{\mu\nu}$
and $T_{\perp r} = l_{\mu}\, z_{r\nu}\, T^{\mu\nu}$ are components
of the energy momentum tensor of the isolated system(see Section III
of Ref.\cite{9}).

\medskip

The Poisson brackets of the non-Grassmann variables are: $\{
z^{\mu}(\tau,\vec{\sigma}), \rho_{\nu}(\tau, {\vec \sigma}_1) \} = -
\delta^{\mu}_{\nu}\, \delta^3(\vec \sigma - {\vec \sigma}_1)$, $\{
A_A(\tau,\vec{\sigma}), \pi^B(\tau, {\vec \sigma}_1) \} =
\delta^B_A\, \delta^3(\vec \sigma - {\vec \sigma}_1)$, $\{
\eta^r(\tau), \kappa_s(\tau) \} = - \delta^r_s$, $\{
\lambda_1(\tau), \pi_{\lambda_1}(\tau) \} = 1$.

\medskip

The Dirac Hamiltonian containing the canonical Hamiltonian
(\ref{2.9}) and the primary constraints is

\bea
  H_D &=& - \lambda_1(\tau)\, \Big(\alpha^*(\tau)\, \alpha(\tau) +
 \beta^*(\tau)\, \beta(\tau)\Big) + \gamma_1(\tau)\, \pi_{\lambda_1}(\tau)
 +\nonumber \\
 &+&\zeta_\mu(\tau)\, \chi^\mu(\tau) + \zeta_{(\alpha)}(\tau)\, \chi_{(\alpha)}(\tau)
 + \zeta_{(\beta)}(\tau)\, \chi_{(\beta)}(\tau) + \zeta_{(\alpha)}^*(\tau)\,
 \chi^*_{(\alpha)}(\tau) + \zeta_{(\beta)}^*(\tau)\,
 \chi^*_{(\beta)}(\tau) +\nonumber \\
 &+& \int d^3\sigma\, \Big[\lambda^{\mu}(\tau, \vec \sigma)\, {\cal
 H}_{\mu}(\tau, \vec \sigma) + \lambda_{\tau}(\tau, \vec \sigma)\,
 \pi^{\tau}(\tau, \vec \sigma) - A_{\tau}(\tau, \vec
 \sigma)\, \partial_r\, \pi^r(\tau, \vec \sigma)\Big],
 \label{3.4}
 \eea

\noindent where $\gamma_1(\tau)$, $\zeta_{\mu}(\tau)$,
$\zeta_{(\alpha)}(\tau)$, $\zeta_{(\beta)}(\tau)$,
$\zeta^*_{(\alpha)}(\tau)$, $\zeta^*_{(\beta)}(\tau)$,
$\lambda^{\mu}(\tau, \vec \sigma)$, $\lambda_{\tau}(\tau, \vec
\sigma)$, are Dirac multipliers.
\medskip

The preservation in time of the primary constraints implies the
following secondary constraints

\bea
 &&\alpha^*(\tau)\, \alpha(\tau) + \beta^*(\tau)\, \beta(\tau) \approx 0,
 \nonumber \\
 && \Gamma(\tau,\vec{\sigma}) = \partial_r\, \pi^r(\tau,\vec{\sigma})
 \approx 0.
 \label{3.5}
 \eea
\medskip
While the Grassmann constraints (\ref{3.1}) are second class, all
the other constraints are first class \footnote{As in Ref.\cite{11}
this is a non trivial check. What turns out to be first class are
not the constraints ${\cal H}_{\mu}(\tau, \vec \sigma) \approx 0$,
but modified constraints obtained by adding to ${\cal H}_{\mu}(\tau,
\vec \sigma)$ linear combinations of the Grassmann constraints
$\chi^{\mu}(\tau) \approx 0$ of Eqs.(\ref{3.1}). Since the
calculations are the same given in Ref.\cite{11}, we do not
reproduce them.} and the variables $\lambda_1(\tau)$, $z^{\mu}(\tau,
\vec \sigma)$, $A_{\tau}(\tau, \vec \sigma)$ and the longitudinal
component of the vector potential are gauge variables.

\medskip

As shown in Ref.\cite{11} we can eliminate the second class
Grassmann constraints (\ref{3.1}) by replacing the Poisson brackets
$\{ \xi^{\mu}(\tau), \pi_{\xi}^{\nu}(\tau) \} = - \eta^{\mu\nu}$,
$\{ \alpha(\tau), \pi_{\alpha}(\tau) \} = \{ \alpha^*(\tau),
\pi^*_{\alpha}(\tau) \} = \{ \beta(\tau), \pi_{\beta}(\tau) \} = \{
\beta^*(\tau), \pi^*_{\beta}(\tau) \} = - 1$ with the following
Dirac brackets (still denoted $\{ .,. \}$ for the sake of
simplicity)

\bea
 &&\{ \xi^{\mu}(\tau), \xi^{\nu}(\tau) \} = - i\,
 \eta^{\mu\nu},\nonumber \\
 &&{}\nonumber \\
 &&\{ \alpha(\tau), \alpha^*(\tau) \} = \{ \beta(\tau),
 \beta^*(\tau) \} = - i.
 \label{3.6}
 \eea

\bigskip

Then the Dirac Hamiltonian becomes

\bea
  H_D &=& - \lambda_1(\tau)\, \Big(\alpha^*(\tau)\, \alpha(\tau) +
 \beta^*(\tau)\, \beta(\tau)\Big) + \gamma_1(\tau)\, \pi_{\lambda_1}(\tau)
 +\nonumber \\
 &+& \int d^3\sigma\, \Big[\lambda^{\mu}(\tau, \vec \sigma)\, {\cal
 H}_{\mu}(\tau, \vec \sigma) + \lambda_{\tau}(\tau, \vec \sigma)\,
 \pi^{\tau}(\tau, \vec \sigma) - A_{\tau}(\tau, \vec
 \sigma)\, \partial_r\, \pi^r(\tau, \vec \sigma)\Big].
 \label{3.7}
 \eea

\subsection{The Transversality Constraint}

At this point we add by hand the transversality constraints on the
Grassmann variables $\xi^{\mu}(\tau)$

\beq
 \Phi(\tau) = P_{\mu}\, \xi^{\mu}(\tau) \approx 0,
 \label{3.8}
 \eeq

\noindent with the conserved total $P^{\mu}$ given in
Eq.(\ref{2.10}). This constraint is second class and eliminates the
time-like component of $\xi^{\mu}_{\tau}$, so that there is only an
independent Wigner spin-1 Grassmann 3-vector

\beq
 \xi^r_{\perp}(\tau) = \epsilon^r_{\mu}(\vec h)\, \xi^{\mu}(\tau).
 \label{3.9}
 \eeq

If we go to Dirac brackets by eliminating the second class
constraint (\ref{3.8})  (so that $\xi^{\mu}(\tau) \equiv
\epsilon^{\mu}_r(\vec h)\, \xi^r_{\perp}(\tau)$), we get

\beq
 \{ \xi^r_{\perp}(\tau), \xi^s_{\perp}(\tau) \}^* = i\, \delta^{rs},
 \label{3.10}
 \eeq

\noindent with all the other basic Poisson brackets left unmodified
except the following ones

\beq
 \{ z^{\mu}(\tau, \vec \sigma), z^{\nu}(\tau, {\vec \sigma}_1) \}^*
 \not= 0,\qquad \{ z^{\mu}(\tau, \vec \sigma), \xi^r_{\perp}(\tau) \}^*
 \not= 0.
 \label{3.11}
 \eeq

\medskip

The new canonical variable ${\tilde z}^{\mu}(\tau, \vec \sigma)$
satisfying the standard Poisson brackets with all the other
variables is

\beq
 {\tilde z}^{\mu}(\tau, \vec \sigma) = z^{\mu}(\tau, \vec \sigma) +
 {i\over 2}\, \epsilon^A_{\nu}(\vec h)\,
 \eta_{AB}\, {{\partial\, \epsilon^B_{\rho}(\vec h)}
 \over {\partial\, P_{\mu}}}\, \epsilon^{\rho}_r(\vec h)\,
 \epsilon^{\nu}_s(\vec h)\, \xi^r_{\perp}(\tau)\,
 \xi^s_{\perp}(\tau).
 \label{3.12}
 \eeq

 \medskip

As a consequence the Lorentz generators of Eqs.(\ref{2.10}) become

\bea
 J^{\mu\nu} &=& \int d^3\sigma\, \Big[{\tilde z}^\mu(\tau,\vec{\sigma})\,
 \rho^\nu(\tau,\vec{\sigma}) - {\tilde z}^\nu(\tau,\vec{\sigma})\,
 \rho^\mu(\tau,\vec{\sigma})\Big] + {\tilde S}^{\mu\nu}_{\xi},
 \nonumber \\
 &&{}\nonumber \\
 &&{\tilde S}^{ij}_{\xi} = \epsilon^{ijr}\, S^r_{\xi}, \qquad
 S^r_{\xi} = - {i\over 2}\, \epsilon^{ruv}\, \xi^u_{\perp}\,
 \xi^v_{\perp}, \nonumber \\
 &&{\tilde S}_{\xi}^{oi} = - {{\epsilon^{ijr}\, P^j\, S^r_{\xi}}\over
 {P^o + \sqrt{\sgn\, P^2}}}.
 \label{3.13}
 \eea

\subsection{The Restriction to the Wigner 3-spaces of the Inertial Rest Frame}

As shown in Refs.\cite{4,9} the restriction to the inertial rest
frame is obtained by adding the gauge fixing $z^{\mu}(\tau, \vec
\sigma) \approx z^{\mu}_{(W)}(\tau, \vec \sigma)$ with the embedding
of Eq.(\ref{1.1}), in which $x^{\mu}_o$ is identified with
$Y^{\mu}(0)$ so that the observer, origin of the 3-coordinates on
the Wigner 3-spaces, becomes the Fokker-Pryce center of inertia,
i.e. $Y^{\mu}(\tau) = z^{\mu}_W(\tau, 0) = Y^{\mu}(0) + h^{\mu}\,
\tau$ ($h^{\mu} = P^{\mu}/\sqrt{\sgn\, P^2} = l^{\mu}$ is the unit
normal to Wigner 3-spaces). Then Eq.(\ref{3.12}) becomes

\bea
 {\tilde z}^{\mu}(\tau, \vec \sigma) &\approx& Y^{\mu}(\tau) +
 \epsilon^{\mu}_r(\vec h)\, \sigma^r + {i\over 2}\,
 \epsilon^A_{\nu}(\vec h)\,
 \eta_{AB}\, {{\partial\, \epsilon^B_{\rho}(\vec h)}
 \over {\partial\, P_{\mu}}}\, \epsilon^{\rho}_r(\vec h)\,
 \epsilon^{\nu}_s(\vec h)\, \xi^r_{\perp}(\tau)\,
 \xi^s_{\perp}(\tau) =\nonumber \\
 &{\buildrel {def}\over =}& {\hat x}^{\mu}(\tau)
 + \epsilon^{\mu}_r(\vec h)\, \sigma^r.
 \label{3.14}
 \eea

\noindent  The new position ${\hat x}^{\mu}(\tau)$ allows the use
Eqs.(\ref{3.13}) to define the following spin-like tensor $S^{\mu\nu}_s =
J^{\mu\nu} - \Big({\hat x}^{\mu}(\tau)\, P^{\nu} - {\hat
x}^{\nu}(\tau)\, P^{\mu}\Big) $.

\medskip

As shown in Ref.\cite{9}, if we evaluate the Poincar\'e generators
(\ref{3.13}) with the embedding $z^{\mu}_{(W)}(\tau, \vec \sigma)$,
we get the following description of the isolated system "atom plus
electro-magnetic field":\medskip

1) There is a decoupled (non-local) non-covariant canonical external
center of mass ${\tilde x}^{\mu}(\tau)$ whose conjugate canonical
momentum is $P^{\mu} = M c\, h^{\mu}$ ($M c = \sqrt{\sgn\, P^2}$ is
the invariant mass of the isolated system)

\bea
 {\tilde x}^{\mu}(\tau) &=& Y^{\mu}(\tau) +
 {i\over 2}\, \epsilon^A_{\nu}(\vec h)\,
 \eta_{AB}\, {{\partial\, \epsilon^B_{\rho}(\vec h)}
 \over {\partial\, P_{\mu}}}\, \epsilon^{\rho}_r(\vec h)\,
 \epsilon^{\nu}_s(\vec h)\, \xi^r_{\perp}(\tau)\,
 \xi^s_{\perp}(\tau) -\nonumber \\
 &-& \epsilon^A_{\nu}(\vec h)\, \eta_{AB}\, {{\partial\,
 \epsilon^B_{\rho}(\vec h)}\over {\partial\, P_{\mu}}}\,
 S^{\mu\nu}_s,\nonumber \\
 &&{}\nonumber \\
 &&\{  {\tilde x}^{\mu}(\tau), P^{\nu} \}^* = - \eta^{\mu\nu},\qquad
 \{ {\tilde x}^{\mu}(\tau), {\tilde x}^{\nu}(\tau) \}^* = 0.
 \label{3.15}
 \eea

This decoupled non-covariant point particle carries a pole-dipole
structure (the invariant mass $M$ and the rest spin $\vec S$ of the
isolated system) and an {\it external} realization of the Poincar\'e
generators \footnote{As shown in Refs.\cite{4,9,15} the
non-covariant canonical center of mass ${\tilde x}^{\mu}(\tau)$ and
its momentum $P^{\mu}$ can be replaced with the frozen Jacobi data
$\vec z$, $\vec h$, $\{ z^i, h^j \} = \delta^{ij}$, by means of the
expressions ${\tilde x}^{\mu}(\tau) = \Big(\sqrt{1 + {\vec h}^2}\,
\Big[\tau + {{\vec h \cdot \vec z}\over {M c}}\Big]; {{\vec z}\over
{M c}} + (\tau + {{\vec h \cdot \vec z}\over {M c}})\, \vec h\Big)$,
$P^{\mu} = Mc\, h^{\mu} = Mc\, \Big(\sqrt{1 + {\vec h}^2}; \vec
h\Big)$. The Cauchy data for the Newton-Wigner position are $\vec
z/Mc$. Then we get that $L^{\mu\nu} = {\tilde x}^{\mu}\, P^{\nu} -
{\tilde x}^{\nu}\, P^{\mu}$ has the components $L^{oi} = - \sqrt{1 +
{\vec h}^2}\, z^i$ and $L^{ij} = z^i\, h^j - z^j\, h^i$.}

\bea
 P^{\mu} &=& M c\, h^{\mu},\qquad J^{\mu\nu} = {\tilde x}^{\mu}\,
 P^{\nu} - {\tilde x}^{\nu}\, P^{\mu} + {\tilde S}^{\mu\nu},\nonumber \\
 &&{}\nonumber \\
 &&{\tilde S}^{\mu\nu} = {\tilde S}_s^{\mu\nu} + {\tilde
 S}_{\xi}^{\mu\nu},\nonumber \\
 &&{\tilde S}^{oi} = {\tilde S}^{oi}_s = - {{\epsilon^{ijr}\, h^j\, S^r}\over
 {1 + \sqrt{1 + {\vec h}^2}}},\nonumber \\
 &&{\tilde S}^{ij} = {\tilde S}^{ij}_s + {\tilde S}^{ij}_{\xi} =
 \epsilon^{ijr}\, S^r.
 \label{3.16}
 \eea

\medskip

2)  Inside the Wigner 3-spaces of the rest frame the system "atom
plus electro-magnetic field" is described by the atom canonical
3-coordinates $\vec \eta(\tau)$, $\vec \kappa(\tau)$ (plus the
Grassmann variables $\xi^r_{\perp}(\tau)$, $\alpha(\tau)$,
$\alpha^*(\tau)$, $\beta(\tau)$, $\beta^*(\tau)$) and by the
canonical coordinates $A_A(\tau, \vec \sigma)$, $\pi^A(\tau, \vec
\sigma)$ of the electro-magnetic field. There is a unfaithful
internal representation representation $M c$, ${\cal P}^r$, $S^r$,
${\cal K}^r$, of the Poincar\'e generators restricted by the
conditions ${\cal P}^r \approx 0$ and ${\cal K}^r \approx 0$ (they
are the rest-frame conditions eliminating the internal center of
mass inside the Wigner 3-space). These generators have the following
expression

\bea
 M c &=& \sqrt{m^2\, c^2 + 2\, mc\, \Omega\, \beta^*(\tau)\, \beta(\tau) +
 \vec{\kappa}^2(\tau)} +\nonumber \\
 &+& \frac{mc}{\sqrt{m^2\, c^2 + \vec{\kappa}^2(\tau)}}\,
 \Big(\beta^*(\tau)\, \alpha(\tau) + \alpha^*(\tau)\, \beta(\tau)\Big)\,
 \vec{d}(\tau) \cdot \vec{\pi}(\tau, \vec \eta(\tau)) +\nonumber \\
 &+&{1\over 2}\, \int d^3\sigma\, \Big(\vec{\pi}^2(\tau,\vec{\sigma}) +
 \vec{B}^2(\tau,\vec{\sigma})\Big), \nonumber \\
 &&{}\nonumber \\
 \vec S &=& \vec \eta(\tau) \times \vec \kappa(\tau) - {i\over 2}\,
 {\vec \xi}_{\perp}(\tau) \times {\vec \xi}_{\perp}(\tau) + \int
 d^3\sigma\, \vec \sigma \times \Big(\vec \pi(\tau, \vec \sigma) \times
 \vec B(\tau, \vec \sigma)\Big),\nonumber \\
 &&{}\nonumber \\
 {\vec {\cal P}} &=& \vec \kappa(\tau) + \int d^3\sigma\, \vec \pi(\tau,
 \vec \sigma) \times \vec B(\tau, \vec \sigma)  \approx 0,\nonumber \\
 {\vec {\cal K}} &=& - \vec \eta(\tau)\,
 \Big(\sqrt{m^2\, c^2 + 2\, mc\, \Omega\, \beta^*(\tau)\, \beta(\tau) +
 \vec{\kappa}^2(\tau)} +\nonumber \\
 &+& \frac{mc}{\sqrt{m^2\, c^2 + \vec{\kappa}^2(\tau)}}\,
 \Big(\beta^*(\tau)\, \alpha(\tau) + \alpha^*(\tau)\, \beta(\tau)\Big)\,
 \vec{d}(\tau) \cdot \vec{\pi}(\tau, \vec \eta(\tau)) \Big) -\nonumber \\
 &-& {1\over {2 c}}\, \int d^3\sigma\, \vec \sigma\,
 \Big({\vec \pi}^2(\tau, \vec \sigma) + {\vec B}^2(\tau,
 \vec \sigma)\Big)\approx 0.
 \label{3.17}
 \eea

The invariant mass $M c = \sqrt{\sgn\, P^2}$ is derived from Eqs.
(\ref{2.10}) and (\ref{3.3}) restricted to $z^{\mu}_{(W)}(\tau, \vec
\sigma)$ by noting that Eq.(\ref{2.6}) and $z^{\mu}_{(W)\, A}(\tau,
\vec \sigma) = \epsilon^{\mu}_A(\vec h)$ imply $A^{\mu\nu}(\tau)\,
\epsilon_{u\mu}(\vec h)\, \epsilon_{v\nu}(\vec h) = 0$ and
$A^{\mu\nu}(\tau)\, h_{\mu}\, \epsilon_{s\nu}(\vec h) =
\Big(\beta^*(\tau)\, \alpha(\tau) + \alpha^*(\tau)\,
\beta(\tau)\Big)\, d^s(\tau)$, where we introduced the electric
dipole

\beq
 \vec d(\tau) = - i\, d\, {\vec \xi}_{\perp}(\tau) \times {\vec
 \xi}_{\perp}(\tau).
 \label{3.18}
 \eeq

\medskip

The non-relativistic limit of the atom energy is $M c^2 = m c^2 +
{{{\vec k}^2(\tau)}\over {2 m}} + {{\Omega}\over m}\,
\beta^*(\tau)\, \beta(\tau) + \Big(\beta^*(\tau)\, \alpha(\tau) +
\alpha^*(\tau)\, \beta(\tau)\Big)\, c\, \vec{d}(\tau) \cdot
\vec{\pi}(\tau, \vec \eta(\tau))$.  After a quantization done with
the method explained in the next Section it reproduces the
Hamiltonian of the Rabi model \cite{1a,2,3} with $\tilde \Omega =
\Omega / m$ and electric dipole $c\, \vec d(\tau)$. By making the
rotating phase approximation one finds the Hamiltonian of the
Jaynes-Cummings model \cite{1,2,3}.

\bigskip

As shown in Refs.\cite{4,9} the Dirac Hamiltonian in the rest frame
is

\bea
 H_D &=& M c - \lambda_1(\tau)\, \Big(\alpha^*(\tau)\, \alpha(\tau) +
 \beta^*(\tau)\, \beta(\tau)\Big) + \gamma_1(\tau)\, \pi_{\lambda_1}(\tau)
 +\nonumber \\
 &+& \int d^3\sigma\, \Big[ \lambda_{\tau}(\tau, \vec \sigma)\,
 \pi^{\tau}(\tau, \vec \sigma) - A_{\tau}(\tau, \vec
 \sigma)\, \partial_r\, \pi^r(\tau, \vec \sigma)\Big].
 \label{3.19}
 \eea

To the resulting Hamilton equations one must add  the rest-frame
constraints ${\vec {\cal P}} \approx 0$ and ${\vec {\cal K}} \approx
0$. With the gauge fixing $\lambda_1(\tau) \approx 0$ we can also
eliminate the gauge variable $\lambda_1(\tau)$: the Dirac
Hamiltonian reduces to $H_D = M c + \int d^3\sigma\, \Big[
\lambda_{\tau}(\tau, \vec \sigma)\, \pi^{\tau}(\tau, \vec \sigma) -
A_{\tau}(\tau, \vec \sigma)\, \partial_r\, \pi^r(\tau, \vec
\sigma)\Big]$ and we have to add the constraint $\alpha^*(\tau)\,
\alpha(\tau) + \beta^*(\tau)\, \beta(\tau) \approx 0$ to the
Hamilton equations.

\subsection{The Restriction to the Radiation Gauge}

In Refs.\cite{4,9} it is shown how to make the restriction to the
radiation gauge, where the electro-magnetic field is described by
the transverse quantities ${\vec A}_{\perp}(\tau, \vec \sigma)$ and
${\vec \pi}_{\perp}(\tau, \vec \sigma)$. In the rest frame we have
${\vec \pi}_{\perp}(\tau, \vec \sigma) = {\vec E}_{\perp}(\tau, \vec
\sigma)$ and $\vec B(\tau, \vec \sigma) = \vec \partial \times {\vec
A}_{\perp}(\tau, \vec \sigma)$. All the previous formulas remain
valid with the replacement $\vec \pi \mapsto {\vec \pi}_{\perp}$.
Now the Dirac Hamiltonian (\ref{3.19}) is $H_D = M c$.

\medskip

As shown in Ref.\cite{4} we have the following representation of the
electro-magnetic fields in the radiation gauge

\bea
 {\vec {\tilde A}}_{\perp}(\tau ,\vec k) &=& {\frac{i}{{{\vec
 k}^2}}}\, \vec k \times {\vec {\tilde B}}(\tau ,\vec k) =
 {\frac{1}{{2\, |\vec k|}}}\, \Big[\vec \alpha (\tau ,\vec k) +
 {\vec \alpha}^*(\tau ,- \vec k)\Big],  \nonumber \\
 {\vec A}_{\perp}(\tau ,\vec \sigma )&=& {\frac{1}{{(2\pi)^3}}}\,
 \int {\frac{ {d^3k}}{{2\, \omega (\vec k)}}}\, \sum_{\lambda =
 1,2}\, {\vec \epsilon} _{\lambda}(\vec k)\, \Big[a_{em\,
 \lambda}(\tau ,\vec k)\, e^{i\, \vec k \cdot \vec \sigma} +
 a^*_{em\, \lambda}(\tau ,\vec k)\, e^{- i\, \vec k
 \cdot \vec \sigma}\Big],  \nonumber \\
 &&{}  \nonumber \\
 &&{}  \nonumber \\
 {\vec \pi}_{\perp}(\tau ,\vec \sigma ) &=& {\frac{i}{{2\,
 (2\pi)^3}}}\, \int d^3k\, \sum_{\lambda = 1,2}\, {\vec
 \epsilon}_{\lambda}(\vec k)\, \Big[ a_{em\, \lambda}(\tau ,\vec k)\,
 e^{i\, \vec k \cdot \vec \sigma} - a^*_{em\, \lambda}(\tau ,\vec
 k)\, e^{- i\, \vec k \cdot \vec \sigma}\Big] =
 \nonumber \\
 &{{\buildrel \circ \over {=}}}_{dyn}& - {\frac{{\partial\, {\vec A}
 _{\perp}(\tau ,\vec \sigma )}}{{\partial\, \tau}}},  \nonumber \\
 &&{}  \nonumber \\
 \vec B(\tau ,\vec \sigma ) &=& {\frac{i}{{2\, (2\pi)^3}}}\, \int
 {\frac{{d^3k }}{{\omega (\vec k)}}}\, \vec k \times \sum_{\lambda =
 1,2}\, {\vec \epsilon} _{\lambda}(\vec k)\, \Big[a_{em\,
 \lambda}(\tau ,\vec k)\, e^{i\, \vec k \cdot \vec \sigma} -
 a^*_{em\, \lambda}(\tau ,\vec k)\, e^{- i\, \vec k
 \cdot \vec \sigma}\Big],  \nonumber \\
 &&{}  \nonumber \\
 &&{}  \nonumber \\
 a_{em\, \lambda}(\tau ,\vec k) &=& \int d^3\sigma\, {\vec \epsilon}
 _{\lambda}(\vec k)\cdot \Big[\omega (\vec k)\, {\vec A}_{\perp}(\tau
 ,\vec \sigma ) - i\, {\vec \pi}_{\perp}(\tau ,\vec \sigma )\Big]\,
 e^{- i\, \vec k \cdot \vec \sigma},\nonumber \\
 &&{}\nonumber \\
 &&\{ a_{em\, \lambda}(\tau
 ,\vec k), a^*_{em\, \lambda^{^{\prime}}}(\tau , {\vec
 k}^{^{\prime}}) \} = - i\, \Omega (\vec k)\, c\, \delta_{\lambda
 \lambda^{^{\prime}}}\, \delta^3(\vec k - {\vec k} ^{^{\prime}}).
 \nonumber \\
 &&{}
 \label{3.20}
 \eea

\vfill\eject

\section{The Quantization}

In Refs.\cite{4,5} we presented the classical theory underlying
relativistic atomic physics. In Ref.\cite{14} we developed a new
version of relativistic quantum mechanics in the inertial rest-frame
instant form consistent with what is known about relativistic bound
states and taking into account the non-covariance of the canonical
relativistic external center of mass. It can be used to quantize the
atoms in the absence of the electro-magnetic field. The main
complication is the imposition of the rest-frame conditions ${\vec
{\cal P}} \approx 0$ and ${\vec {\cal K}} \approx 0$ due to the
complicated form of the boost generators in presence of interactions
among the particles. When they cannot be solved at the classical
level (which would allow quantization of only physical degrees of freedom),
one must quantize all the canonical variables in the Wigner 3-space
in a un-physical Hilbert space and then select the physical states
by asking that they satisfy $< \Phi_{phys} | {\hat {\vec {\cal P}}}
| \Phi_{phys} > = 0$ and $< \Phi_{phys} | {\hat {\vec {\cal K}}} |
\Phi_{phys} > = 0$. This type of quantization has not yet been done
for the free transverse electro-magnetic field in the radiation
gauge much less for the system of charged particles plus a generic
transverse electro-magnetic field.

\medskip

Therefore we will give a quantization of the atom in an external
classical transverse electro-magnetic field and then we will show
its coupling to a single mode of a free field (this is suitable for
a two-level atom) as it is often done in atomic physics \cite{2}.

\bigskip

The canonical 3-coordinates $\vec \eta(\tau)$, $\vec \kappa(\tau)$,
are quantized in the standard way: in the coordinate representation
we have $\vec \eta \mapsto \vec \eta$ and $\vec \kappa \mapsto -i
\hbar\, {{\partial}\over {\partial\, \vec \eta}}$ as unbounded
operators in a Hilbert space with the standard scalar product.

\medskip

The Grassmann variables $\xi^r_{\perp}$ describing the electric
dipole are quantized to the Pauli matrices, $\xi^r_{\perp} \mapsto
\sqrt{{{\hbar}\over 2}}\, \sigma^r$ like for the spinning particle
of Ref.\cite{11}. Therefore, Eqs.(\ref{3.18}) gives a quantum
electric dipole ${\hat {\vec d}}$ behaving like a spin 1/2 Pauli
particle.

\medskip

Each pair of complex  Grassmann variables $\alpha$, $\alpha^*$, and
$\beta$, $\beta^*$ is quantized to operators $\hat a$, ${\hat
a}^{\dagger}$, and $\hat b$, ${\hat b}^{\dagger}$ corresponding to
Fermi oscillators, namely satisfying the anti-commutation relations
$[\hat a, {\hat a}^{\dagger} ]_+ = [\hat b, {\hat b}^{\dagger} ]_+ =
\hbar$ and with one Fermi oscillator commuting with the other one.
Therefore we get a 4-dimensional Hilbert space whose states have the
form

\bea
 \Psi &=& C_{++}\, \Psi_\alpha(+) \otimes \Psi_\beta(+) + C_{-+}\,
 \Psi_\alpha(-) \otimes \Psi_\beta(+) +\nonumber \\
   &+& C_{+-}\, \Psi_\alpha(+) \otimes \Psi_\beta(-) + C_{--}\,
   \Psi_\alpha(-) \otimes \Psi_\beta(-),\nonumber \\
   &&{}\nonumber \\
   &&\hat a\, \Psi = C_{-+}\, \Psi_\alpha(+) \otimes \Psi_\beta(+) +
   C_{--}\, \Psi_\alpha(+) \otimes \Psi_\beta(-),\nonumber \\
   &&{\hat a}^{\dagger}\, \Psi = C_{++}\, \Psi_\alpha(-) \otimes
   \Psi_\beta(+) + C_{+-}\, \Psi_\alpha(-) \otimes
   \Psi_\beta(-),\nonumber \\
   &&\hat b\, \Psi = C_{++}\, \Psi_\alpha(+) \otimes \Psi_\beta(-) +
   C_{-+}\, \Psi_\alpha(-) \otimes \Psi_\beta(-),\nonumber \\
   &&{\hat b}^{\dagger}\, \Psi = C_{+-}\, \Psi_\alpha(+) \otimes \Psi_\beta(+) +
   C_{--}\, \Psi_\alpha(-) \otimes \Psi_\beta(+),\nonumber \\
   &&{}
 \label{4.1}
 \eea

If we choose the following ordering

 \bea
 \alpha^*\, \alpha &\mapsto& - {\hat a}^{\dagger}\, \hat a +\frac{\hbar}{2},
 \nonumber \\
 &&{}\nonumber \\
 \beta^*\, \beta &\mapsto& {\hat b}^{\dagger}\, \hat b -
 \frac{\hbar}{2},
 \label{4.2}
 \eea

\noindent then the constraint $\alpha^*(\tau)\, \alpha(\tau) +
\beta^*(\tau)\, \beta(\tau) \approx 0$ implies the following
condition on the physical states

\bea
 &&\Big({\hat b}^{\dagger}\, \hat b - {\hat a}^{\dagger}\, \hat
 a\Big)\, \Psi_{phys} = 0,\nonumber \\
 &&{}\nonumber \\
 &&\Downarrow\nonumber \\
 &&{}\nonumber \\
 &&\Psi_{phys} = C_{-+}\, \Psi_\alpha(-) \otimes \Psi_\beta(+) +
 C_{+-}\, \Psi_\alpha(+) \otimes \Psi_\beta(-) =\nonumber \\
 &{\buildrel {def}\over =}& C_{-+}\, \Phi(+) +
 C_{+-}\, \Phi(-).
  \label{4.3}
  \eea

If we define the operators

\beq
 \hat c = {\hat b}^{\dagger}\, \hat a, \qquad {\hat c}^{\dagger} =
 \hat b\, {\hat a}^{\dagger},
 \label{4.4}
 \eeq

\noindent we get

\bea
 &&\hat c \, \Phi(+) = \Phi(-),\qquad \hat c\, \Phi(-) = 0,
 \nonumber \\
 &&{\hat c}^{\dagger}\, \Phi(-) = \Phi(+),\qquad
 {\hat c}^{\dagger}\, \Phi(+) = 0,\nonumber \\
 &&{}\nonumber \\
 &&{\hat b}^{\dagger}\, \hat b\, \Psi_{phys} =
 {\hat c}^{\dagger}\, \hat c\, \Psi_{phys}.
 \label{4.5}
 \eea

Therefore the mass term $mc + \Omega\, \beta^*\, \beta$ will become
the mass $mc + \Omega\, {\hat c}^{\dagger}\, \hat c$ of the two
physical levels.

\medskip

Moreover Eq.(\ref{4.4}) implies that the coupling term of  the
electric dipole to the electric field present in $M c$ of
Eq.(\ref{3.17}) takes the form

\beq
 {{m c}\over {\sqrt{m^2\, c^2 + {\hat {\vec \kappa}}^2}}}\,
 \Big({\hat c}^{\dagger} + \hat c\Big)\, {\hat {\vec d}}
 \cdot {\vec E}(\tau, \vec \eta(\tau)).
 \label{4.6}
 \eeq

\bigskip

As is usually done in atomic physics, in particular in the treatment
of two-level atoms \cite{2,3}, let us consider only one quantized
mode with energy $\hbar\, \omega\, ({\hat a}^{\dagger}_{em}\, {\hat
a}_{em} + {1\over 2})$ of the transverse electro-magnetic field
(\ref{3.20}). Then the transverse electric field can be approximated
as ${\vec {\cal E}}(\tau)\, ({\hat a}_{em} + {\hat
a}^{\dagger}_{em})$.

\medskip

Since in the physical Hilbert space the operators ${\hat
c}^{\dagger}\, \hat c$, ${\hat c}^{\dagger}$ and $\hat c$ coincide
with the Pauli matrices $\sigma_3$, $\sigma_+$ and $\sigma_{-}$
respectively, in the radiation gauge the quantum Dirac Hamiltonian
takes the following form

\bea
 {\hat H}_D &=& \sqrt{m^2\, c^2 + 2\, mc\, \Omega\, \sigma_3 +
 {\hat {\vec k}}^2} + \hbar\, \omega\, \Big({\hat a}^{\dagger}_{em}\, {\hat
 a}_{em} + {1\over 2}\Big) +\nonumber \\
 &+&  {{m c}\over {\sqrt{m^2\, c^2 + {\hat {\vec \kappa}}^2}}}\,
  \Big({\hat {\vec d}} \cdot {\vec {\cal E}}(\tau)\Big)
 \Big({\hat c}^{\dagger} + \hat c\Big)\, ({\hat a}_{em} + {\hat
 a}^{\dagger}_{em}),
 \label{4.7}
 \eea

\noindent whose non-relativistic limit is ($\tilde \Omega =
\Omega/m$)

\bea
 \hat H &=& {{{\hat {\vec \kappa}}^2}\over {2m}} +
 {{\hbar}\over 2}\, \tilde \Omega\, \sigma_3 +
 \hbar\, \omega\, \Big({\hat a}^{\dagger}_{em}\, {\hat
 a}_{em} + {1\over 2}\Big) +\nonumber \\
 &+&\Big(c\, {\hat {\vec d}} \cdot {\vec {\cal E}}(\tau)\Big)\,
 \Big(\sigma_+ + \sigma_{-}\Big)\, ({\hat a}_{em} + {\hat
 a}^{\dagger}_{em}).
 \label{4.8}
 \eea

But this is this is the Hamiltonian of the Rabi model \cite{1a,2,3},
which becomes the Jaynes-Cummings one \cite{1,2,3} in the rotating
phase approximation, in which the last term becomes $\Big(c\, {\hat
{\vec d}} \cdot {\vec {\cal E}}(\tau)\Big)\,
 \Big(\sigma_+\, {\hat a}_{em} + \sigma_{-}\, {\hat
 a}^{\dagger}_{em}\Big)$.

\medskip

In the rest frame description one should add the rest-frame
conditions $< \Phi_{phys} | {\hat {\vec {\cal P}}} | \Phi_{phys} > =
0$ and $< \Phi_{phys} | {\hat {\vec {\cal K}}} | \Phi_{phys} > = 0$
in the same approximation.

\vfill\eject

\section{Conclusions}

Our model for the neutral two-level atom allowed us to get the
relativistic generalization of the Hamiltonians of the Rabi and
Jaynes-Cummings models. Moreover we found the explicit form of the
Poincar\'e generators. We also showed how one could describe a
charged two-level atom with also a magnetic dipole. These
relativistic models could be applied in future use of two-level
atoms in space experiments near the Earth, where atomic physics must
take into account both special relativity and post-Newtonian general
relativity.

\bigskip

To arrive at these results we had to describe the isolated system
atom plus dynamical electro-magnetic field as a parametrized
Minkowski theory. Instead usually in atomic physics one considers
the two-level atom interacting with an external electro-magnetic
field in the absolute Euclidean 3-space of Galilei space-time
identified with the Euclidean 3-space of a relativistic inertial
frame. But this is  relativistically problematic because interaction
terms like $\vec d \cdot \vec E$ should be interpreted in the
inertial frame instantaneously comoving with the atom, which has an
accelerated motion in 3-space. Moreover the transversality
constraint (\ref{3.8}) is not well defined if $P^{\mu}$ is the
non-conserved momentum of the atom. All these problems are solved by
our approach.

\bigskip

Finally let us point out the main open problem in the
description of extended atoms. If we have an extended object as an
atom with N constituents, we may describe it as a point-like object
by using its multipolar expansion (see Section 5 of Ref.\cite{5})
and replace the equations of motions of the N constituents with the
resulting (often approximate) equations of motions for the
multipoles. However we do not have a consistent variational
principle containing a point particle (the monopole) carrying all
the other multipoles and interacting with the electro-magnetic field
so as to get a consistent set of equations of motion (and conservation
laws) for the multipoles. The solution of this problem would allow us
to describe N-level atoms with arbitrary multipoles.

\vfill\eject


\begin{thebibliography}{}

 \bibitem{1a}I.I.Rabi, {\it On the Process of Space Quantization},
 Phys.Rev. {\bf 49}, 324 (1926); {\it Space Quantization in a Gyrating
 Magnetic Field}, {\bf 51}, 652 (1937).


\bibitem{1}E.T.Jaynes and F.W.Cummings, {\it Comparison of Quantum
and Semi-Classical Radiation Theories with Application to the Beam
Maser}, Proc. JEEE {\bf 51}, 89 (1963).\hfill\break
 B.W.Shore and P.L.Knight, {\it The Jaynes-Cummings Model},
 J.Mod.Opt. {\bf 40}, 1195 (1993).



\bibitem{2} W.P.Schleich, \textit{Quantum Optics in Phase Space} (Wiley-VCH,
Berlin, 2001).


\bibitem{3}J.Larson, {\it Dynamics of the Jaynes-Cummings and Rabi
Models: Old Wine in New Bottles}, Phys. Scr. {\bf 76}, 146 (2007)
(quant-ph/0612095).







\bibitem{4}D.Alba, H.W.Crater and L.Lusanna, {\it Towards Relativistic
Atom Physics. I. The Rest-Frame Instant Form of Dynamics and a
Canonical Transformation for a system of Charged Particles plus the
Electro-Magnetic Field}, Canad.J.Phys. {\bf 88}, 379 (2010) (arXiv:
0806.2383).

\bibitem{5}D.Alba, H.W.Crater and L.Lusanna, {\it Towards Relativistic
Atom Physics. II. Collective and Relative Relativistic Variables for
a System of Charged Particles plus the Electro-Magnetic Field},
Canad.J.Phys. {\bf 88}, 425 (2010) (arXiv:0811.0715).


\bibitem{6} D.Alba, H.W.Crater and L.Lusanna, \textit{Hamiltonian
Relativistic Two-Body Problem: Center of Mass and Orbit
Reconstruction}, J.Phys. {\bf A40}, 9585 (2007) (gr-qc/0610200).


\bibitem{7} D.Alba, H.W.Crater and L.Lusanna, \textit{The Semiclassical
Relativistic Darwin Potential for Spinning Particles in the Rest
Frame Instant Form: Two-Body Bound States with Spin 1/2
Constituents}, Int.J.Mod.Phys. \textbf{A16}, 3365-3478 (2001)
(hep-th/0103109).


\bibitem{8}L.Lusanna, {\it The N- and 1-Time Classical
Descriptions of N-Body Relativistic Kinematics and the
Electromagnetic Interaction}, Int.J.Mod.Phys. {\bf A12}, 645 (1997).
\hfill\break L.Lusanna, {\it The Chrono-Geometrical Structure of
Special and General Relativity: A Re-Visitation of Canonical
Geometrodynamics}, lectures at 42nd Karpacz Winter School of
Theoretical Physics: Current Mathematical Topics in Gravitation and
Cosmology, Ladek, Poland, 6-11 Feb 2006, Int.J.Geom.Methods in
Mod.Phys. {\bf 4}, 79 (2007). (gr-qc/0604120).




\bibitem{9}D.Alba and L.Lusanna,
{\it Charged Particles and the Electro-Magnetic Field in
Non-Inertial Frames: I. Admissible 3+1 Splittings of Minkowski
Spacetime and the Non-Inertial Rest Frames},  Int.J.Geom.Methods in
Physics {\bf 7}, 33 (2010) (0908.0213)


\bibitem{10} D.Alba and L.Lusanna,
{\it Charged Particles and the Electro-Magnetic Field in
Non-Inertial Frames: II. Applications: Rotating Frames, Sagnac
Effect, Faraday Rotation, Wrap-up Effect }, Int.J.Geom.Methods in
Physics {\bf 7}, 185 (2010) (0908.0215).

\bibitem{11} F.Bigazzi and L.Lusanna, {\it Spinning Particles on
Spacelike Hypersurfaces and their Rest Frame Description},
Int.J.Mod.Phys. {\bf A14}, 1429 (1999) (hep-th/9807052).

\bibitem{12}D.Alba,H.W. Crater  and L.Lusanna,
{\it Massless Particles plus Matter in the Rest-Frame Instant Form
of Dynamics}, {\it J.Phys.} A {\bf 43} 405203  (arXiv 1005.5521).

\bibitem{13}D.Alba, H.W.Crater  and
L.Lusanna,  {\it The Rest-Frame Instant Form and Dirac Observables
for the Open Nambu String}, Eur.Phys.J.Plus {\bf 126}, 26 (2011)
(arXiv 1005.3653).


\bibitem{14}D.Alba,H.W. Crater  and L.Lusanna, {\it  Relativistic
Quantum Mechanics and Relativistic Entanglement in the Rest-Frame
Instant Form of Dynamics}, to appear in J.Math.Phys.  (arXiv
0907.1816).



\bibitem{15} D.Alba and L.Lusanna, \textit{Quantum Mechanics in Noninertial
Frames with a Multitemporal Quantization Scheme: I. Relativistic
Particles}, Int.J.Mod.Phys. \textbf{A21}, 2781 (2006)
(hep-th/0502194).

\bibitem{16} D.Alba, \textit{Quantum Mechanics in Noninertial Frames with a
Multitemporal Quantization Scheme: II. Nonrelativistic Particles},
Int.J.Mod.Phys. \textbf{A21}, 3917 (hep-th/0504060).

\bibitem{17} M. O. Scully, and M. S. Zubairy, Quantum optics, (Cambridge,
1997)












\end{thebibliography}
\end{document}